\documentclass[referee]{raa}           
\usepackage{graphicx,times}
\usepackage{natbib}
\usepackage{amssymb,amsmath}
\usepackage{float}
\usepackage{amsmath} 
\usepackage[colorlinks=true,linkcolor=blue,citecolor=blue]{hyperref}
\usepackage{color, colortbl}
\bibpunct{(}{)}{;}{a}{}{,}

\begin{document}

   \title{Day-timescale Quasi-periodic Oscillations of the Gev BL Lac RX J0805.4+7534 with TESS}

 \volnopage{ {\bf 20XX} Vol.\ {\bf X} No. {\bf XX}, 000--000}
   \setcounter{page}{1}

   \author{Xin-Shun Jin 
   \inst{1,2}, Ting-Feng Yi\inst{1,2,*}, Yang-Wei Zhang\inst{3,*}, Yun-Cai Shen\inst{1,2}, Jun-Jie Wang
      \inst{1,2}, Li-Sheng Mao\inst{1,2}, Liang Dong\inst{2,4}
   }
%% Here is an example of three authors come from different institutes.
%% For single author or all the authors from an institute, use "\inst{}" only
\institute{Key Laboratory of Colleges and Universities in Yunnan Province for High-energy Astrophysics, Department of Physics, Yunnan Normal University, Kunming 650500, China.\\
    \and
        Yunnan Province China-Malaysia HF-VHF Advanced Radio Astronomy Technology International Joint Laboratory, Kunming 650011, China.
    \and
        School of Astronomy and Space Sciences, University of Chinese Academy of Sciences, 100049, Beijing, China.\\
    \and 
        Yunnan Observatories, Chinese Academy of Sciences, 396 Yangfangwang, Guandu District, Kunming, 650216, People’s Republic of China.\\
    \and
        Email: yitingfeng@ynnu.edu.cn; zhangyangwei@ucas.ac.cn\\
\vs \no
    {\small Received 20XX Month Day; accepted 20XX Month Day}}
\abstract{This paper reports for the first time the detection of quasi-periodic oscillations (QPOs) in the light curves of the BL Lacertae object RX J0805.4+7534. The Transiting Exoplanetary Survey Satellite (TESS) observed this source in seven Sectors of the sky, and we extracted the light curves for these Sectors using a custom method. The presence of QPO signals was found in these light curves. To detect the periodicity and assess the statistical significance of the QPO signals, we employed two methods: Lomb-Scargle periodograms and Weighted Wavelet \textit{Z}-transform. Both of these different methods yielded consistent results. The results show that QPO signals exist in Sectors 20, 26, and 53, with the confidence levels exceeding 99.73\%. The QPOs in Sectors 20 and 26 are $\sim 3.8$ days and have a global significance of 95\%. To explain these rapid quasi-periodic variations, we discussed several possible physical scenarios. The most possible one is the kink instability in relativistic jets. The other possible scenario is the rotation of hot-spots in the innermost accretion disk. Based on the hot-spot orbital model hypothesis, the mass of the black hole at the center of this BL Lac object ($4.09 \times 10^9$ $M_\odot$ for the maximum Kerr black hole) was estimated. The validity of these two explanations requires further observational data to verify. However, since the radiation of BL Lac objects primarily originates from jets, we prefer the kink instability in relativistic jets to be the cause of this rapid QPO.
\keywords{BL Lacertae objects --- RX J0805.4+7534 --- Optical light curve --- Black hole}
}

   \authorrunning{X.-S. Jin et al. }            %author_head in even pages
   \titlerunning{Optical Quasi-periodic Oscillations of the Gev BL Lac RX J0805.4+7534 with TESS}  % title_head in odd pages
   \maketitle
%________________________________________________ sections below
%

\section{Introduction}           %% first-level sections will be auto-capitalized
\label{sect:intro}
The center of an active galactic nucleus (AGN) is a supermassive black hole surrounded by an accretion disk and hot gas, which produce radiation across the entire electromagnetic spectrum. The energy originates from the accretion of gas and matter onto the black hole, with the associated radiation coming from the accretion disk and relativistic jets \citep{2017A&ARv..25....2P}. Depending on the orientation of the relativistic jets relative to the observer, AGNs can be classified as quasars, radio galaxies, or blazars. Blazars as an extreme subclass of AGNs, exhibit relativistic jets that are oriented nearly directly towards Earth. This results in the concentration and enhancement of non-thermal radiation from the relativistic jets, making this radiation dominant across the entire electromagnetic spectrum \citep{1995PASP..107..803U, 1993ApJ...407...65G}. Based on their spectral characteristics and observational properties, blazars are classified into two main types: BL Lac objects and flat-spectrum radio quasars (FSRQ). Based on synchronous peak frequencies, BL Lac can be further divided into three types: low-frequency synchronous peak BL Lac (peak frequency less than $10^{14}$ Hz, LBL), medium-frequency synchronous peak BL Lac (peak frequency greater than $10^{14}$ Hz and less than $10^{15}$ Hz, MBL) and high-frequency synchronous peak BL Lac (peak frequency greater than $10^{15}$ Hz, HBL) \citep{2010ApJ...716...30A}. Observations and studies indicate that BL Lac exhibit significant variations in their flux on timescales ranging from minutes to years (see \citealp{1993ApJ...411..614U, 2004A&A...419..485C, 2008ApJ...689...79C, 2011ApJ...730L...8A, 2014A&A...562A..79S, 2019ApJ...887..185S, 2020ApJ...891..120B, 2020PASP..132d4101Y, 2024MNRAS.531.3927M}, and references therein). 

In addition to stochastic variability, among the population of BL Lac, only a handful of objects exhibit periodic variations in their light curves, and one of these types of periodic variability is quasi-periodic oscillation (QPOs) \citep{10.1093/mnras/sty2720}. This scarcity is partially due to irregular sampling and instrumental limitations, resulting in limited ground-based detection of such data. However, besides the red noise introduced by observational effects such as irregular sampling, these periodic signals may also be obscured by intrinsic stochastic red noise, leading to ambiguity in their identification \citep{2016MNRAS.461.3145V}. Although QPOs are typically transient in nature and have an extremely low probability of appearing in light curves, studying them is of great significance for our deeper understanding of the internal mechanisms, core structure, and physical properties of BL Lac objects \citep{2019MNRAS.482.1270C, 2020ApJS..250....1T, 2021MNRAS.506.1540L}. The QPOs in BL Lac exhibit many types across timescales, ranging from minutes to days, months, or even years. Different models have been used to explain the physical mechanisms of QPOs on different scales (see \citealp{2009ApJ...690..216G, 2015ApJ...813L..41A, 2018NatCo...9.4599Z, 2020A&A...642A.129S, 2022ApJ...931..168G, 2023MNRAS.522L..84K},and references therein). With the development of observational techniques, an increasing number of day-timescale QPO signals from BL Lac and specifically HBL have been detected in the optical band, below are 11 BL Lac that exhibit QPOs with periods on the day-timescale, i.e., S5 0716+714 \citep{2018AJ....155...31H}; PKS 1440--389 \citep{2024Univ...10..242L}; S4 0954+658 \citep{2023ApJ...943...53K}; BL Lacaertae, 1RXS J004519.6+212735, 1RXS J111741.0+254858  \citep{2024MNRAS.527.9132T}; QSO J2345--1555, QSO B0422+004, ATPMN J090453.4--573503, 1RXS J002159.2--514028, QSO B0537--441 \citep{2024MNRAS.528.6608T}. However, another subclass of blazars, FSRQs, is rarely reported to exhibit optical day-timescale QPO signals. The recent literature provides only one case, namely S5 1044+71 \citep{2025RAA....25b5004W}. The ratio of this kind of FSRQs to BL Lac is about 9\%. Based on this roughly statistical analysis, we found that BL Lac seem to have a higher probability of having a QPO of several days in the optical band than other types of AGNs. Whether this empirical rule holds true requires further data to support it.

Searching for optical day-timescale QPOs in BL Lac requires observational data with an extremely high cadence. However, ground-based observations are constrained by a variety of factors, such as, irregular sampling, limited cadence, and short temporal baselines \citep{2016MNRAS.461.3145V}. These limitations make it difficult to reliably identify day-timescale QPOs and to assess their statistical significance. Therefore, high-cadence and uniformly sampled long-term photometric observations are essential for further investigating the prevalence and properties of day-timescale QPOs in BL Lac objects. The observational data from the Transiting Exoplanet Survey Satellite (TESS)  \citep{2015JATIS...1a4003R} seems meets these requirements. The TESS has largely overcome the limitations of ground-based observations \citep{2015JATIS...1a4003R}, providing regularly sampled light curve data. In this work, we downloaded all light curve data for BL Lac objects from the TESS database. Through observation and analysis, we discovered a new QPO signal in the BL Lac object RX J0805.4+7534. This was first observed by the ROSAT All-Sky Survey, and \cite{1996A&A...309..419N} identified it as a BL Lac type object. Its redshift is z = 0.121 \citep{2000ApJS..129..547B}, and its peak frequency is greater than $10^{15}$ Hz, classifying it as a HBL object, and belong to GeV Fermi blazars \citep{2017ApJS..232...18A}. With the development of observation technology, RX J0805.4+7534 has been observed in multiple spectral bands. Furthermore, the source was identified as a TeV candidate \citep{2021ApJ...916...93Z}. 

This paper reports for the first time the discovery of a QPO signal in the light curve of the BL Lac RX J0805.4+7534. The paper is structured as follows: Section \ref{sec2} briefly introduces the basic information and instrument details of the TESS, as well as the observation results and data processing of RX J0805.4+7534. Section \ref{sec3} briefly introduces various analytical techniques used to verify the existence of the QPO signal and presents the analytical results. Section \ref{sec4} presents the conclusions and discussion.

\section{TESS Observations and Data Reduction} \label{sec2}
\subsection{TESS Observations}
TESS is a space telescope that is primarily  used to track highly luminous and precise stars across almost the entire sky in search of transiting exoplanets. The instrument is equipped with four wide-field optical charge-coupled device (CCD) cameras, each with a 24°×24° field of view. These four cameras are linked together to image a 24° × 96° area of the sky. In its early stages (2018-2019), TESS was observed at a 30-minute interval \citep{2015JATIS...1a4003R}, but now (2020-) at a 10-minute or 2-minute interval. TESS divides the sky into 26 observation Sectors, 13 in each hemisphere, and completes one sector in approximately 27 days. Therefore, completing observations over all sky regions would require approximately two years. Observations would first be carried out in the Southern Hemisphere, followed by those in the Northern Hemisphere using the same observational strategy, with each hemisphere observed for one year \citep{2015JATIS...1a4003R}.

\subsection{Data Reduction}
We used the \textbf{Quaver code} \footnote{\url{https://github.com/kristalynnesmith/quaver}} to extract all light curve data for BL Lac and Seyfert galaxies from the TESS official website \footnote{\url{https://archive.stsci.edu/tess/bulk_downloads/bulk_downloads_ffi-tp-lc-dv.html}}. This code is a customized extraction and correction tool specifically designed for TESS AGN light curves. Its primary advantage lies in the ability for users to customize the extraction region directly from full-frame images, thereby eliminating the need to download entire images. This extraction method reduces interference from neighbouring sources \citep{2023ApJ...958..188S}. It is worth noting that this code is publicly available and comes with detailed documentation.

The three analysis methods used in this work—the PCA-hybrid method, the simple hybrid method, and the fully hybrid method—are effective in mitigating systematic effects in the TESS light curve data \citep{2023ApJ...958..188S}. Each method exhibits distinct strengths and limitations. The PCA method is straightforward to implement, requiring only the specification of principal component parameters. Its limitations include potential biases when correcting instrumental systematic errors and limited adaptability to long-term light variations. The simple hybrid method extends PCA by incorporating additive background effects and systematic error corrections, thereby preserving long-term variability. The simple hybrid method, while producing light curves that closely resemble those obtained from ground-based synchronous observations, suffers from a lower sampling density compared to ground-based data. The fully hybrid method further accounts for all remaining systematic errors beyond those addressed by the previous two methods, thereby effectively mitigating systematic effects. However, it may overfit long-term variability in the light curves (see \citealp{2023ApJ...958..188S, 2024ApJ...966..158P, 2025MNRAS.tmp.1810T}). In our study, we selected the fully hybrid method photocurve for analysis.

\begin{figure*}[ht!]
    \centering
    \includegraphics[width=0.8\textwidth]{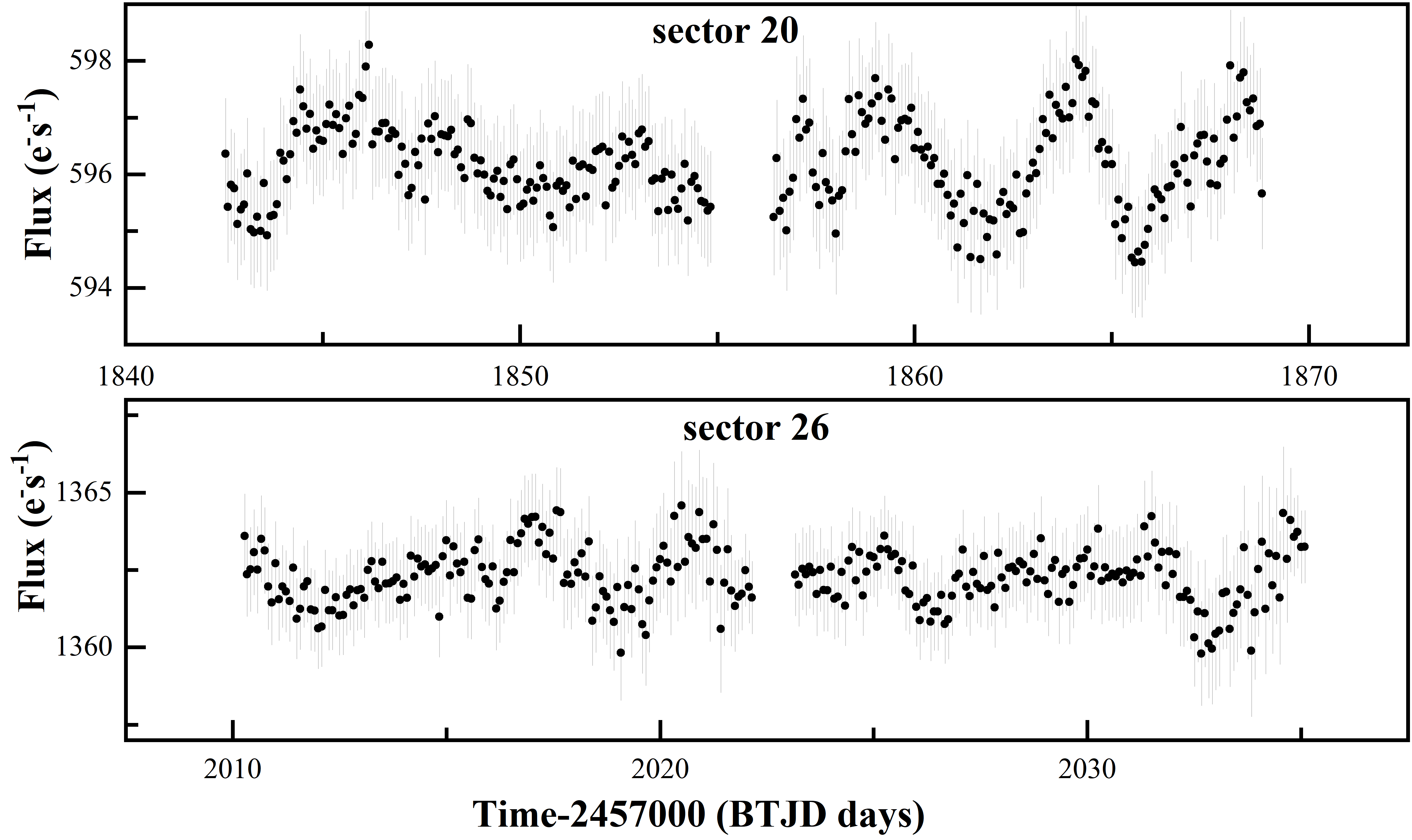}
    \caption{The light curves of the BL Lac RX J0805.4+7534 in Sectors 20 (top panel) and 26 (bottom panel) (2 hr bin). Black dots represent flux values, while gray curves indicate the error.
    \label{fig:fig1}}
\end{figure*}

\begin{figure*}[ht!]
    \centering
    \includegraphics[width=1\textwidth]{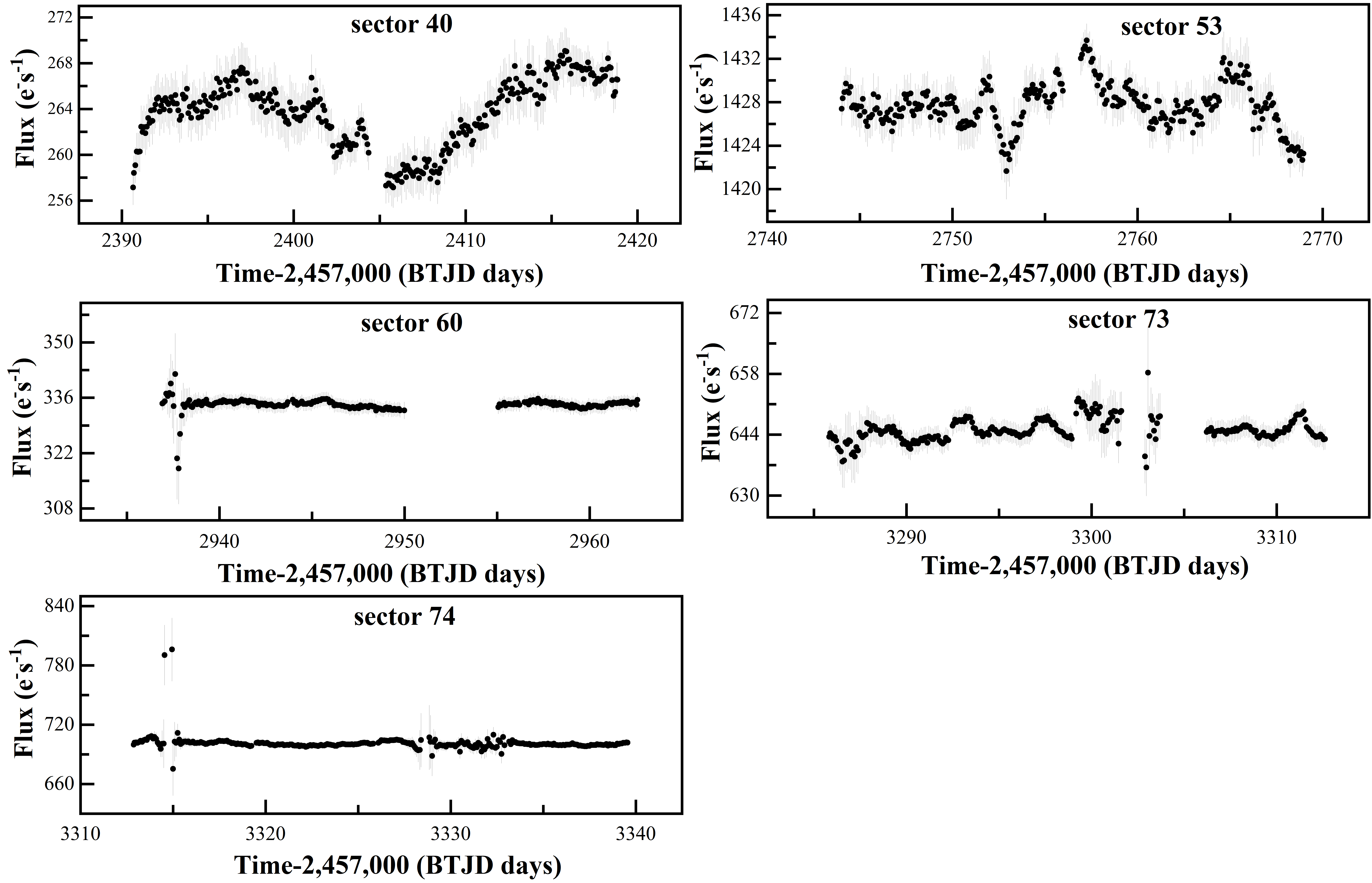}
    \caption{The light curves of the BL Lac RX J0805.4+7534 in Sectors 40, 53, 60, 73 and 74 (2 hr bin). Black dots represent flux measurements, while gray curves indicate the errors.
    \label{fig:fig2}}
\end{figure*}

This work used a fully hybrid method to correct data from all Sectors of RX J0805.4+7534. Given the large number of data points per sector, we averaged the data using a 2 hr bin scheme, this processing has little effect on the overall light curve and can be neglected. Figures \ref{fig:fig1} and \ref{fig:fig2} show the light curves of the seven Sectors. TESS observed this source in Sector 26 on December 24, 2019, concluding in Sector 74 on January 4, 2024. The observations were intermittent and occurred in Sectors 26, 40, 60, 53, and 73. We have noted gaps of approximately one or two days in the light curves for these Sectors. This irregularity may be caused by the satellite transmitting data to Earth or waiting for download commands during this period, resulting in irregular data sampling.

\section{Data Analysis} \label{sec3}
 we employ two methods—the Lomb–Scargle Periodogram() and the weighted wavelet \textit{z}-transform—to detect potential QPO signals in RX J0805.4+7534 and estimate their significance using a Monte Carlo approach. The following subsections provide a brief introduction to these methods and present the results of their application.

\subsection{Lomb–Scargle Periodogram (LSP)}
LSP \citep{1976Ap&SS..39..447L, 1982ApJ...263..835S} is a classic algorithm for extracting periodic signals from non-uniformly sampled time series and use the $\chi^2$ statistic to fit the data of the entire time series \citep{2023ApJ...943...53K}. It performs trigonometric function-related summations and operations on non-uniformly sampled time series to quantify the signal power at different frequencies $\omega$, thereby determining the presence of periodic signals in the time series \citep{2018ApJS..236...16V}. The expression for LSP is as follows:
\begin{equation}
P_Y(2\pi f) = \frac{\left[ {\sum_{i=1}^{n}} Y_i \cos(2\pi f)(t_i - \tau) \right]^2}{2 {\sum_{i=1}^{n}} \cos^2(2\pi f)(t_i - \tau)} + \frac{\left[ {\sum_{i=1}^{n}} Y_i \sin(2\pi f)(t_i - \tau) \right]^2}{2 {\sum_{i=1}^{n}} \sin^2(2\pi f)(t_i - \tau)},
\end{equation}
among these, $P_Y(2\pi f)$ is the power spectral density at frequency $f$; $n$ indicates the number of data points; $Y_i$ is the observation value at the $i$ sampling point; $t_i$ is the time at the isampling point, where the time phase correction $\tau$ is defined as
\begin{equation}
\tau = \frac{1}{4\pi f} \arctan\left[ \frac{\sum_{i=1}^{n} \sin 4\pi f t_i}{\sum_{i=1}^{n} \cos 4\pi f t_i} \right].
\end{equation}
In the evaluation of LSP, in order to avoid the overestimation of LSP power due to oversampling, the frequency range is generally as follows: the maximum frequency is taken as the Nyquist frequency $f_{\rm Nyq} = \frac{1}{2\langle \Delta t \rangle}$, minimum frequency $ f_{min} = \frac{1}{T}$, the corresponding frequency point $f(i) = \frac{i}{T}, \quad i = 1, 2, 3...$, where $T$ represents the length of the time series \citep{2005A&A...431..391V, 2020PASP..132d4101Y}. To implement this approach, the LSP code provided by PyAstronomy \footnote{\url{https://github.com/sczesla/PyAstronomy}} was employed in this paper \citep{2019ascl.soft06010C}.

\subsection{Weighted Wavelet \textit{Z}-transform (WWZ)}
WWZ is another alternative method for time series analysis, particularly well-suited in astronomy for analyzing irregularly sampled, non-stationary, and noisy data. It can further detect transient periodic signals within such data \citep{1996AJ....112.1709F}. This method convolves the light curve with a time-frequency correlation kernel, and then transforms the data to the time-frequency domain to generate a WWZ plot. Its advantage lies in detecting dominant periodic signals and their temporal duration \citep{2022MNRAS.510.3641R}. The WWZ transform employs the Morlet wavelet as its basis function, with the \textit{Z} variable defined as follows:
\begin{equation}
WWZ = \frac{1}{2} \left[ \frac{N_{\text{eff}} \, V_Y}{V_X - V_Y} - \frac{3V_Y}{V_X - V_Y} \right],
\end{equation}

\begin{equation}
N_{\text{eff}} = \frac{\left[ \sum e^{-2a\omega_0^2 (t_j - \tau_0)^2} \right]^2}{\sum e^{-2a\omega_0^2 (t_j - \tau_0)^2}},
\end{equation}
\begin{equation}
V_{\text{X}}= \frac{\sum_j \omega_j X^2t_j}{\sum_\alpha \omega_\alpha} - \left[ \frac{\sum_j \omega_j Xt_j}{\sum_\alpha \omega_\alpha} \right]^2, V_{\text{Y}}= \frac{\sum_j \omega_j Y^2t_j}{\sum_\alpha \omega_\alpha} - \left[ \frac{\sum_j \omega_j Yt_j}{\sum_\alpha \omega_\alpha} \right]^2,
\end{equation}
where, $N_{eff}$ represents the number of valid data points, $V_X$ denotes the weighted variance of the data function, and $V_Y$ denotes the weighted variance of the simulation function \citep{1996AJ....111..541F, 2024ApJ...961..180L}. In this wavelet transform, the decay constant $c$ should be less than $\frac{1}{8}\pi^{-2}$ \citep{1996AJ....112.1709F}, according to the value of $c$ mentioned in \cite{2023ApJ...943...53K}'s article, when $c = 0.001$, the $e^{-\frac{1}{4c}}$ in the Morlet wavelet $f(\textit{z}) = e^{-c\textit{z}^2} \left( e^{i\textit{z}} - e^{-\frac{1}{4c}} \right)$ is very small and can even be neglected \citep{2026A&A...707A.371G}. To implement this method, we use the Python package \footnote{\url{https://github.com/eaydin/WWZ}} provided by \cite{2017zndo....375648A}. 

\subsection{Light Curve Simulation and Confidence Significance}
Due to the complex internal physical mechanisms of AGNs and various observational and instrumental instabilities during photometric observations, photometric light curves often exhibit red-noise–like behavior, which can lead to spurious QPO detections. Consequently, it is crucial to assess the statistical significance of the detected QPO signal. Given the characteristics of red noise processes in periodograms, it is essential to investigate the effects of non-uniform sampling on periodogram noise \citep{2014MNRAS.445..437M}. Following the methodology described by \cite{2005A&A...431..391V}, we fitted the power spectral density (PSD) of the sector light curves (LCs) using a power-law model (i.e.,$P(f) \propto Af^{-\beta}$), where, $A$ is the normalization constant, \textit{f} is the temporal frequency, $\beta$ is the power spectral index. The low-frequency power excess is consistent with red-noise behavior, while the flattening of the spectrum at high frequencies indicates the dominance of white noise (for example, $\beta \approx 0$ is mainly white noise, and $1 \le \beta \le 2$, generally associated with red noise.) (see, \citealp{2025MNRAS.541.2955P, 2026MNRAS.545f2211T}, and references therein). The PSDs of the Sectors were fitted individually. For Sectors 20, 26, and 53, the $\beta$ values are $1.18\pm0.05$, $0.73\pm0.06$ and $1.50\pm0.059$, respectively. The ratios of $\chi^2$ to degrees of freedom are 0.93, 0.98, and 0.90, respectively. For Sectors 40 and 73, their $\beta$ values are $1.25\pm0.065$ and $0.97\pm0.062$, and the ratios of chi-square to degrees of freedom are 0.91 and 0.73, respectively. Since reduced chi-square values close to unity indicate satisfactory fits, these results demonstrate that the adopted power-law model provides an adequate description of the PSDs in all analyzed Sectors. To evaluate the significance of the periodic signal detected in RX J0805.4+7534, we adopted the algorithm proposed by \citet{2013MNRAS.433..907E} and utilized the “\textbf{DELightcurveSimulation}” \footnote{\url{https://github.com/samconnolly/DELightcurveSimulation}} Python package developed by Connolly to simulate $5\times 10^4$ artificial light curves. These artificial curves share the fundamental properties of the original light curve \citep{2024MNRAS.531.3927M}. Subsequently, we assessed the confidence level for this source by calculating the LSP and WWZ power for each artificial light curve, thereby evaluating the robustness of the detected periodic signal.

In the LSP analysis, the local significance should be corrected for multiple testing effects in order to derive the corresponding global significance, which provides a more robust assessment of the statistical significance of a detected periodic signal. We use the methods described by \cite{2005A&A...431..391V} and \cite{2022ApJ...926L..35O} to calculate the global significance. According to the formula mentioned in \cite{2024ApJ...965..124A}, global significance can be approximated as: $P_{\text{global}} = 1 - (1 - P_{\text{local}})^N$, where$P_{\text{local}}$, means local significance, $N$ is the number of valid experiments, which is the product of the number of sources being searched and the number of independent frequencies in each periodogram, and can be estimated using Monte Carlo \citep{2020A&A...634A.120A, 2025MNRAS.541.2955P}. Considering the frequency range, we limited the frequency range to between the minimum frequency $(1/T)$ and the Nyquist frequency, discarding frequency points corresponding to the white-noise-dominated region \citep{2005A&A...431..391V, 2026ApJ...998..317K}. The global significance calculations were performed only on Sectors 20, 26, and 53; the periodic signals detected in these Sectors all exceeded the 99.73\% local significance threshold. Since no peaks in Sectors 40 and 73 reached this criterion, global significance corrections were not applied to these Sectors.

 \subsection{Results}
\begin{figure*}[ht!]
    \centering
    \includegraphics[width=0.8\textwidth]{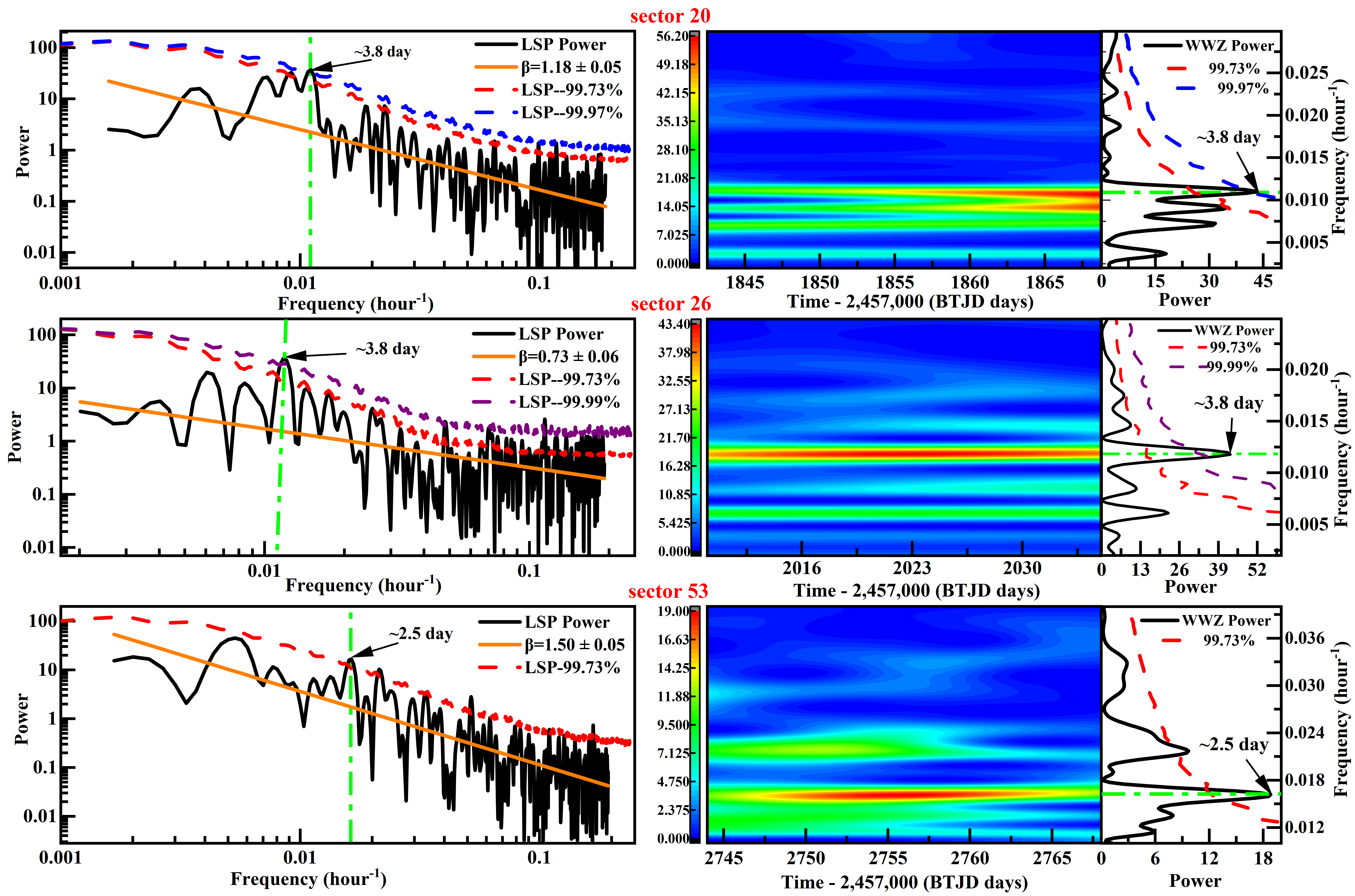}
    \caption{Left: Log-periodogram of the fitted LSP for Sectors 20, 26, and 53; the frequencies in the periodogram range from the minimum frequency to the Nyquist frequency, where the black curve represents the power $\beta$, the orange diagonal line is the power spectral index, the red dash curve shows 99.73\% confidence, blue one represents 99.97\%, and purple one represents 99.99\%. Right: shows the WWZ results for Sectors 20, 26, and 53. In each panel, the black curve represents the time-averaged WWZ power, while the red, blue, and purple dash curves indicate percentile levels corresponding to different confidence levels. A color scale is also included to illustrate the WWZ power distribution over time and frequency. For the WWZ color map, where blue indicates minimum power and red indicates maximum power.
    \label{fig:fig3}}
\end{figure*}

As shown in figures \ref{fig:fig3} and figures \ref{fig:fig4}, we present the LSP and WWZ analysis results of the light curves of five Sectors (Sectors 20, 26, 40, 53, and 73) of RX J0805.4+7534, with the LSP plots using a logarithmic scale. For these light curves, a periodic signal is considered reliable only if a peak exceeds the 99.73\% significance lever in both the LSP and WWZ analyzes.

Figure \ref{fig:fig3} presents the analysis results for RX J0805.4+7534 in Sectors 20, 26 and 53. For sector 20, the LSP reveals a prominent peak at $\sim 3.8$ days, exceeding the 99.97\% local significance level. We take the half width at half-maximum (HWHM) of the peak as the period error (i.e., $3.8\pm0.2$ days). After correcting for multiple trials to obtain the global significance, this peak corresponds to a global significance of roughly 95\%. Wavelet analysis indicates that the $\sim 3.8$ days peak is accompanied by other lower-significance features, with their intensities increasing during the latter half of the observational period. In the time-averaged WWZ map, the $\sim 3.8$ days peak also surpasses the 99.97\% significance level. Taken together, these results suggest that the 3.8 days signal is statistically robust and can be identified as a QPO. For Sector 26, the results of the LSP analysis show a significant peak at $3.8\pm0.4$ days, exceeding the 99.99\% significance level. Our results on correcting for local significance show that 99.99\% corresponds to a global significance of 95\% after correction. The wavelet analysis indicates that the peak at $\sim 3.8$ days is present throughout most of the observational interval, although its power is somewhat reduced during the early and late stages. The signal is most prominent in the middle portion of the dataset. In the time-averaged WWZ spectrum, the $\sim 3.8$ days peak likewise exceeds the 99.99\% significance level. Taken together, the results of both the LSP and WWZ analyses support the interpretation of the $\sim 3.8$ days feature as a candidate QPO signal. For sector 53, the analysis results from LSP and WWZ show that the peaks at about $2.5\pm0.3$ days and $1.9\pm0.1$ days both exceed 99.73\%. However, the WWZ power map shows that the peak at $\sim 2.5$ days is concentrated in the middle portion of the observation and displays stronger power than the other candidate feature. Therefore, the $\sim 2.5$ days signal is favored as the more plausible periodic component. Nevertheless, its global significance does not exceed the 95\% confidence level. In addition, its statistical significance is lower than those of the candidate QPO signals identified in Sectors 20 and 26, rendering the detection less compelling.

Figure \ref{fig:fig4} shows the results of the LSP and WWZ analyses for the light curves of RX J0805.4+7534 in Sectors 40 and 73. No significant periodic signals are detected in either sector, as none of the peaks exceed the 99.73\% significance level in either method. Consequently, these Sectors were not considered further in the QPO analysis, and no global significance correction was applied.
\textcolor{red}{\begin{figure*}[ht!]
    \centering
    \includegraphics[width=0.8\textwidth]{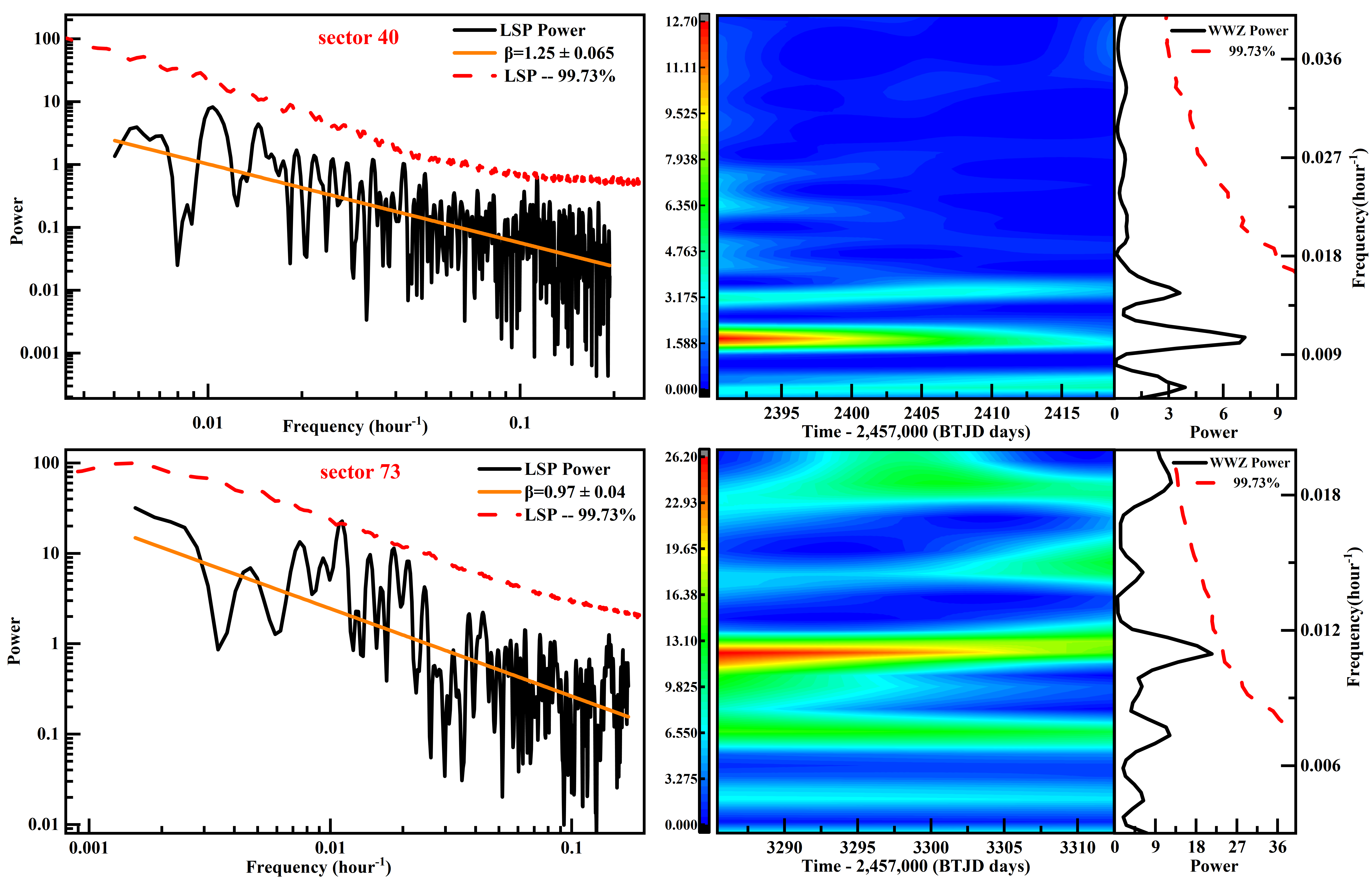}
    \caption{Results for Sectors 40 and 73, with the same description as in figure 3.
    \label{fig:fig4}}
\end{figure*}}

In this study, we selected light curve data from five Sectors, processed them using the two analytical methods described above, and presented the corresponding results. For Sectors 60 and 74, since the overall light curves did not show significant fluctuations, WWZ and LSP were not used to analyze these two Sectors. Furthermore, all light curves analyzed in this paper have gap of approximately 1–2 days, and gaps are  much smaller than the observation duration of the corresponding Sectors and can therefore be ignored \citep{2025RAA....25b5004W, 2024MNRAS.527.9132T}.

\section{Discussion and Conclusion} \label{sec4}
The results show that QPOs may exist in Sectors 20, and 26, with a consistent period of $\sim 3.8$ days and confidence levels exceeding $3.8\sigma$. It seems to be a statistical trend that BL Lac have a higher probability of experiencing QPO in the light band for a few days compared to other types of AGNs. The discovered QPO in BL Lac RX J0805.4+7534 adds a new case to this statistical trend. Its physical origin is currently unclear and requires further investigation.

The radiation mechanisms in blazars are highly dynamic and multifaceted, with non-thermal processes in relativistic jets dominating their radiation output, significantly surpassing the combined radiation from the host galaxy and accretion disk \citep{2001ApJS..134..181J, 2024MNRAS.527.9132T}. This arises from a specific relativistic effect constrained by the observer's angle to the jet ($<10^\circ$) \citep{1996ASPC..110..391U}. The radiation of blazars exhibits strong variability, the time scales can be a few minutes to several years \citep{2010MNRAS.402.2087V, 2014APh....54....1A,2016ApJ...824L..20A, 2017A&A...603A..25A}. However, not all variabilities are interpreted as QPO signals in blazars, only a minority demonstrate statistically significant QPO signals (see, \citealp{1998ApJ...504L..71H, 2013MNRAS.436L.114K, 2015Natur.518...74G, 2001ApJS..134..181J, 2023ApJ...943...53K}, and references therein). These meaningful signals differ in their time scale, radiation mechanisms, and interpretations. For example, the supermassive binary black hole systems \citep{valtonen2008massive} and Lense-Thirring precession of accretion disks \citep{1998ApJ...492L..59S} have been used to explain those periodic signals time scale several years. Visibly, these explanations are not suitable for interpreting the QPO signals with a time scale of several days that we have analyzed here. For these short-term QPOs, with periods on the order of a few days, two possible scenarios have been proposed: one arising from the relativistic jet, and the other originating from the accretion disk. Regarding the quasi-periodic features generated within the jet, a particularly relevant mechanism involves kink instabilities \citep{2020MNRAS.494.1817D,2025ApJ...995...76H}. Kink instability is a current-driven magnetohydrodynamic (MHD) instability that can develop in relativistic jets, and the observed quasi-periodicity may arise as a consequence of this process \citep{2024MNRAS.527.9132T}. Relativistic MHD simulations have shown that kink instabilities are likely to occur in jets dominated by strong toroidal magnetic fields. The growth of the instability can distort the magnetic field configuration and enhance particle acceleration, resulting in the formation of kinked structures that are quasi-periodically distributed along the jet. As these structures evolve and propagate downstream, the associated compression of magnetized plasma regions can modulate the emitted radiation, giving rise to observable QPO signals on timescales of days to weeks (see, \citealp{2009ApJ...700..684M,2017MNRAS.469.4957B, 2024MNRAS.527.9132T, 2024MNRAS.528.6608T}, and references therein). The period of the QPOs is closely related to the growth timescale of the knots, which can be estimated from the lateral motion of the knot regions \citep{2009ApJ...700..684M, 2020MNRAS.494.1817D}. Following \cite{2020MNRAS.494.1817D}, the growth timescale of the kink instability $\tau_{\rm KI}$ can be estimated from the transverse motion of the kink region. Specifically, $\tau_{\rm KI}$ is defined as the ratio of the jet's transverse displacement from its central axis $R_{\rm KI}$ to the average transverse velocity $v_{\rm tr}$, i.e.,
\begin{equation}
 \ T_{\rm obs} = \dfrac{R_{\rm KI}}{\langle \nu_{\rm tr} \rangle \delta}
\end{equation}
where, $\delta$ is the Doppler factor, $R_{\rm KI}$ is the size of a typical flare emission region, approximately $10^{16} - 10^{17}$, and $\nu_{\rm tr}$ is about 0.16c \citep{2020MNRAS.494.1817D}. When $\delta=15$, the observed period $T_{\rm obs}$ is approximately 1-10 days. This timescale is consistent with that of the QPO signals identified in our observational analysis, suggesting that the development of kink instabilities in relativistic jets may be responsible for the observed quasi-periodicities \citep{2024MNRAS.528.6608T}. Alternatively, the observed QPOs may be associated with processes in the accretion disk. Although the emission from blazars is generally dominated by relativistically beamed jet radiation, fluctuations originating in the innermost regions of the accretion disk may modulate the emitted flux, thereby producing QPO signatures with relatively small amplitudes \citep{2020MNRAS.494.1817D, 2023ApJ...943...53K, 2024Univ...10..242L}. Under the assumption that the observed QPOs are produced by hot spots orbiting at the innermost stable region of the accretion disk, the mass of the central black hole can be estimated from the observed QPOs. \citep{1993ApJ...411..602C, 2023MNRAS.520.4118C}. The formula \cite{2009ApJ...690..216G} is:
\begin{equation}
\frac{M_{\rm{BH}}}{M_{\odot}} = \frac{3.23 \times 10^4 P}{\left(r^{3/2} + a\right)(1 + z)}
\end{equation}
where, $M_\odot$ is the solar mass, \textit{P} is the period in seconds, \textit{a} is the angular momentum parameter of the supermassive black hole, \textit{z} is the cosmic redshift of the source (\textit{z} = 0.121), \textit{r} is the ISCO radius in units of $\frac{G M_{\text{BH}}}{c^2}$. Under these conditions, we obtain estimated masses of  $4.09 \times 10^9$ $M_\odot$ for the maximum Kerr black hole ($\textit{a} = 0.9982$, $\textit{r} = 1.2$). However, the estimated black hole mass should be treated with caution and requires further verification through independent dynamical measurements. In comparison, the QPOs detected in this work, with periods of several days, are more plausibly explained by kink instabilities developing within the relativistic jet. Despite extensive studies, the physical origin of QPOs in blazars remains a matter of debate, regardless of whether they occur on year-long or day-long timescales. Therefore, to reliably characterize the behavior of QPOs and understand the physical mechanisms, additional observational data (particularly high-quality polarization measurements) are essential.

\normalem 
\begin{acknowledgements}
This research was supported by the National Natural Science Foundation of China (Grant Nos.: 12203041) and by the Yunnan Province China-Malaysia HF-VHF Advanced Radio Astronomy Technology International Joint Laboratory (Nos. 202303AP140003). This study has made use of the TESS data, obtained from the Mikulski Archive for Space Telescopes (MAST), provided by the NASA Explorer Program.
\end{acknowledgements}

\bibliographystyle{raa}
\bibliography{bibtex}

\end{document}